\documentclass[conference]{IEEEtran}
\IEEEoverridecommandlockouts

\usepackage{cite}
\usepackage{amsmath,amssymb,amsfonts}
\usepackage{algorithmic}
\usepackage{graphicx}
\usepackage{import}
\usepackage{textcomp}
\usepackage{xcolor}
\usepackage{hyperref}
\usepackage[nolist]{acronym}
\def\BibTeX{{\rm B\kern-.05em{\sc i\kern-.025em b}\kern-.08em
    T\kern-.1667em\lower.7ex\hbox{E}\kern-.125emX}}
\begin{document}
\begin{acronym} 
\acro{aas}[AAS]{Asset Administration Shell}
\acro{i40}[I4.0]{Industry~4.0}
\acro{ai}[AI]{Artificial Intelligence} 
\acro{cxe}[cxE]{clipx ENGINEER}
\acro{llm}[LLM]{Large Language Model}
\acro{ann}[ANN]{Approximate Nearest Neighbor}
\acro{cl}[CL]{Category Level}
\acro{bm25}[BM25]{Best Match~25}
\acro{mrr}[MRR]{Mean Reciprocal Rank}
\acro{ir}[IR]{Information Retrieval}
\end{acronym}

\makeatletter 
\newcommand{\linebreakand}{%
  \end{@IEEEauthorhalign}
  \hfill\mbox{}\par
  \mbox{}\hfill\begin{@IEEEauthorhalign}
}
\makeatother 

\title{ECLASS-Augmented Semantic Product Search for Electronic Components
\thanks{This work was supported by the German Federal Ministry of Research, Technology and Space (BMFTR) within the project KI-OWL under grant No. 01IS24057D. It has also been funded by the Phoenix Contact-Stiftung.}
}

\author{
    \IEEEauthorblockN{1\textsuperscript{st} Nico Baumgart}
    \IEEEauthorblockA{\textit{Department of Computer Science and Automation} \\
    \textit{OWL University of Applied Sciences and Arts}\\
    Lemgo, Germany \\
    0009-0000-1390-0794}
    \and
    \IEEEauthorblockN{2\textsuperscript{nd} Markus Lange-Hegermann}
    \IEEEauthorblockA{\textit{Department of Computer Science and Automation} \\ 
    \textit{OWL University of Applied Sciences and Arts}\\
    Lemgo, Germany \\
    0000-0002-5327-4529}
    \linebreakand
    \IEEEauthorblockN{3\textsuperscript{rd} Jan Henze}
    \IEEEauthorblockA{\textit{Data Science and Engineering} \\
    \textit{Phoenix Contact GmbH \& Co. KG}\\
    Blomberg, Germany \\
    jan.henze@phoenixcontact.com}
}

\maketitle

\begin{abstract}
    Efficient semantic access to industrial product data is a key enabler for factory automation and emerging LLM-based agent workflows, where both human engineers and autonomous agents must identify suitable components from highly structured catalogs. However, the vocabulary mismatch between natural-language queries and attribute-centric product descriptions limits the effectiveness of traditional retrieval approaches, e.g., \acs{bm25}. In this work, we present a systematic evaluation of \acs{llm}-assisted dense retrieval for semantic product search on industrial electronic components, and investigate the integration of hierarchical semantics from the ECLASS standard into embedding-based retrieval. Our results show that dense retrieval combined with re-ranking substantially outperforms classical lexical methods and foundation model web-search baselines. In particular, the proposed approach achieves a Hit\_Rate@5 of 94.3\%, compared to 31.4\% for \acs{bm25} on expert queries, while also exceeding foundation model baselines in both effectiveness and efficiency. Furthermore, augmenting product representations with ECLASS semantics yields consistent performance gains across configurations, demonstrating that standardized hierarchical metadata provides a crucial semantic bridge between user intent and sparse product descriptions.
\end{abstract}


\section{Introduction}
\label{introduction}
\begin{figure*}[htbp]
  \centering
  \def\svgwidth{\textwidth}
  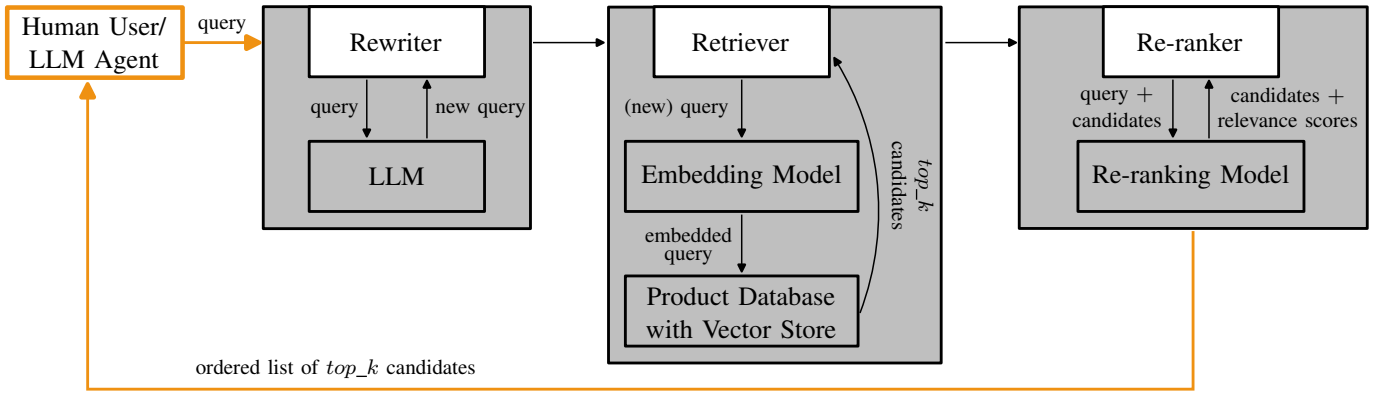
    \caption{Overview of the proposed \ac{llm}-assisted semantic product search pipeline for industrial component retrieval, following the current state of the art in \acl{ir}. A natural-language query from a human user or \ac{llm} agent is optionally rewritten into an attribute-focused representation, embedded into a shared vector space, and used to retrieve top $k$ candidate products via dense similarity search over a vector store. The retrieved candidates are subsequently refined by a re-ranking model that jointly evaluates query-product relevance.}
\label{fig_pipeline}
\end{figure*}
The ongoing fourth industrial revolution, also known as \ac{i40}, is accelerating the digital transformation of manufacturing by integrating technologies such as the Internet of Things, \ac{ai}, and Big Data into production environments. A central concept in this context is the \ac{aas}, which provides a standardized digital representation of industrial assets and enables interoperability across heterogeneous systems \cite{SakuradaTenYears2025}. To ensure semantic interoperability, standardized vocabularies such as ECLASS are widely used to describe products in a machine-interpretable way \cite{ECLASSReferences2026}. In particular, ECLASS is an international classification and description standard for products and services that organizes them in a hierarchical taxonomy and defines shared names, attributes, and semantics~\cite{ECLASSSpecification2026}.

In this setting, efficient access to product information is increasingly relevant for both engineering tasks and \ac{llm}-based agentic workflows. Data representations such as knowledge graphs and embedding-based retrieval systems enable engineers and autonomous agents to query, reason over, and act upon industrial data \cite{ren2025ai}. Nevertheless, identifying suitable components for specific requirements remains difficult due to the highly specialized nature of industrial product data. In contrast to consumer e-commerce catalogs, which often contain rich intent-oriented descriptions and user feedback, such as in the \textit{STaRK-Amazon} benchmark \cite{WuSTaRKBenchmarking2024}, industrial catalogs are typically dominated by technical key-value attributes, categorical fields, and concise formal specifications. This difference creates a persistent gap between natural-language search intent and manufacturer-specific product representations that complicates semantic search and retrieval in industrial contexts.

Traditional lexical retrieval approaches, such as \textit{\ac{bm25}} \cite{RobertsonBM251994}, are computationally efficient but often struggle with vocabulary mismatch, especially when users or \ac{llm} agents are unfamiliar with manufacturer-specific terminology \cite{SongLLMsSparse2025}. Recent advances in \ac{llm}s and dense retrieval have changed retrieval system design by combining vector search with query rewriting and re-ranking (Figure \ref{fig_pipeline}), thereby enabling semantic matching beyond lexical overlap \cite{ZhuLargeLanguage2025}. While these approaches achieve strong results in general domains, their behavior on structured industrial catalogs with attribute-centric product descriptions remains insufficiently studied.

Furthermore, the use of industrial taxonomies as an explicit signal in \ac{ir} remains underexplored. Although such standards have been established to provide formal definitions, unambiguous classifications, and interoperable semantics, current pipelines rarely incorporate their hierarchical metadata into embedding-based search. Leveraging this structured context can improve the disambiguation of specialized technical components from similar distractors.

This paper addresses these gaps by systematically evaluating \ac{llm}-assisted dense retrieval techniques for semantic product search in the industrial domain of control cabinets using $10,346$ products from Phoenix Contact. We investigate how retrieval components, including embedding models, query rewriting, re-ranking, and hyperparameter settings, interact with structured product data. In addition, we integrate hierarchical semantics from the ECLASS standard into embedding-based retrieval and provide a detailed analysis of how different category levels and vocabulary elements influence search performance. The evaluation is conducted on a manually curated dataset that combines expert and non-expert perspectives, enabling both quantitative and qualitative analysis results.

Our findings show that including ECLASS semantics into product embeddings significantly enhances search performance, especially when combined with re-ranking, and outperforms both lexical and foundation model web-search baselines with our best configuration reaching a Hit\_Rate@5 of $94.3\%$ on expert queries. By bridging modern \ac{ai} techniques with standardized industrial semantics, this work contributes to advancing practical applications of semantic search in factory automation and engineering workflows.

\section{Related work}
\label{related_work}
\acp{llm} have been increasingly employed in \ac{i40} to process and retrieve valuable information from unstructured text, such as engineering documents, and answer technical questions about specific domains. For example, Kunze et al. \cite{KunzeAutomatedExtraction2025} utilized \ac{llm}-based agent systems to automatically extract complex conditional causal rules from textual control narratives, thereby reducing manual extraction efforts and supporting root-cause analysis. In other domains, LLMs are leveraged as a chatbot interface for industrial threat models \cite{FregnanLLMsForThreadModels2025}, or the extraction of action sequences from maintenance reports for the efficient maintenance of switchgears \cite{TamascelliLLMsMaintenanceRule2025}, showing the widespread relevance of \acp{llm} across industries.

In the context of the \ac{aas}, \ac{llm} research is primarily focused on the creation of \ac{aas} models and mapping proprietary or manufacturer-specific data to standardized vocabularies, such as ECLASS \cite{AlexopoulosWhyAssetAdministration2025}. For instance, \cite{shi2025dual} introduces a dual data mapping system that leverages fine-tuned \acp{llm} as entity matchers to semantically link proprietary data to \ac{aas} models, and subsequently match \ac{aas} properties to ECLASS dictionary entries. Similarly, both \cite{XiaGenerationAssetAdministration2024} and \cite{shi2024interoperable} use embedding-based semantic matching to map heterogeneous data to the ECLASS description standard. Beyond standardization, generalized embedding models specifically designed for \ac{i40} capture the semantic relationships of industrial asset operations, assisting in tasks like finding relevant sensors for specific equipment failure modes \cite{ConstantinidesGeneralizedEmbedding2025}. 

While the aforementioned embedding-based approaches consistently use the retrieve-and-rerank paradigm, which is a common design pattern in modern \ac{ir} systems \cite{ZhuLargeLanguage2025}, the creation of retrieval pipelines for industrial product search requires application-specific design choices. For instance, \cite{SongLLMsSparse2025} focuses on the storage overhead and the lack of interpretability of dense retrieval approaches by proposing \ac{llm}-based sparse retrievers for the first retrieval stage reaching but not exceeding the performance of dense methods in their experiments. Contrarily, \cite{FreymuthHierarchicalMultiField2025} deals with the challenge of heterogeneous user queries that vary substantially in specificity and level of detail. The authors create field-level product representations by capturing dependencies within an ordered hierarchy of fields. These hierarchical representations enable multi-level retrieval, where an aggregated representation handles efficient initial shortlisting, and field-level vectors dynamically match queries of varying complexity. The latter is technically most similar to our approach, however, the type of data and our goal differ significantly. While Freymuth et al. \cite{FreymuthHierarchicalMultiField2025} use a self-defined hierarchy of standard e-commerce product fields ordered by the average length, we use the formal, strictly defined industrial taxonomy provided by the ECLASS standard. Furthermore, we aim to investigate the benefit of additional ECLASS vocabulary such as category definitions and keywords instead of learning useful product representations.

To conclude, the potential of \ac{llm}-assisted dense retrieval for semantic product search in industrial contexts using established international description standards is underexplored. While there is a growing body of work on \ac{llm} applications in \ac{i40}, the specific challenges of semantic search over structured industrial catalogs and the integration of formal taxonomies like ECLASS into dense retrieval pipelines have not been systematically studied. This paper addresses this gap by evaluating how modern retrieval techniques perform on industrial product data and how leveraging ECLASS semantics can enhance search performance.

\section{Methodology}
\label{methodology}

\subsection{ECLASS product database}
\label{database}
The product database used in this work is based on ECLASS~13.0, an international classification and description standard that provides a structured framework for product categorization. ECLASS defines a hierarchical classification scheme consisting of the four levels \textit{segment}, \textit{main group}, \textit{group}, and \textit{commodity class} and each product is assigned to exactly one commodity class (Figure \ref{fig_database}). In this hierarchy, every category is associated with a unique id, a name, a textual definition, and keywords describing the products belonging to that category. Furthermore, ECLASS specifies both a \textit{basic} and an \textit{advanced} attribute set for each commodity class, where \textit{basic} represents a subset of the \textit{advanced} attributes focusing on the most relevant properties. For further details, the reader is referred to the official technical specification \cite{ECLASSSpecification2026}.

In practice, ECLASS is recommended by the \textit{Plattform Industrie 4.0} as the preferred dictionary of standardized semantics for the definition of digital twins and \ac{aas} \cite{belyaev2021modelling}, which is why it is widely adopted across industries \cite{ECLASSReferences2026}. Hence, there already exists a vast amount of products that potentially benefit from our ECLASS-augmented semantic product search. To ensure a manageable evaluation, we focus on a representative subset of products from the domain of control cabinet components. Specifically, we use all products available in the \ac{cxe} software by Phoenix Contact \cite{PhoenixContactcxE2026}, resulting in a database of $10,346$ products from $122$ commodity classes.

From this database, we construct three variants with different levels of information, which we refer to as \textit{minimal}, \textit{basic}, and \textit{advanced} product data levels. While \textit{advanced} includes all attributes of a product that are defined in the official ECLASS~13.0 \textit{advanced} specification including a manufacturer-specific short description, our \textit{basic} product data level follows the ECLASS~13.0 \textit{basic} specification but also includes the short description, as this is the only textual information beyond key-value pairs. Additionally, \textit{minimal} only consists of the product name, article number, and the short description, to evaluate how the amount and type of available product information influence search performance.

\subsection{Product search pipeline}
\label{common_techniques}
Analogous to the recent trend of \ac{llm}-assisted \ac{ir} systems~\cite{ZhuLargeLanguage2025}, we adopt a three-stage semantic search pipeline consisting of a query rewriter, a retriever, and a re-ranker, as depicted in Figure \ref{fig_pipeline}. We perform ablation studies on each component and analyze the impact of core hyperparameters.

\textbf{Rewriting:}
The query rewriter is responsible for reformulating the user's query into a more effective form for retrieval. Since we focus on dense retrieval, i.e. vector similarity search, in which the query and the product information are embedded into the same vector space, this step aims to align their structure and terminology to facilitate similarity search. Given that our product data primarily consists of structured key-value pairs, we employ an \ac{llm} to transform natural-language queries into concise, attribute-focused expressions. For instance, the query \textit{"Which terminal blocks can I use for KNX applications?"} is rewritten to \textit{"KNX terminal blocks"} eliminating filler words that are not contained in the product descriptions. We evaluate the effectiveness of this step by comparing performance with and without query rewriting.

\textbf{Retrieval:}
In dense retrieval, the product information of each product is embedded into a vector space in advance using an embedding model. At query time, the (rewritten) query is also embedded and compared to the product embeddings to retrieve the most similar products as candidates. In our experiments, we use cosine similarity as the similarity measure, which is a common choice in the literature \cite{ZhaoDenseText2024}. This approach introduces several hyperparameters that can impact search performance, including the embedding size and the number of candidates retrieved based on vector similarity, which we refer to as $top\_k$.

The embedding size refers to the dimensionality of the vector space in which the products and queries are embedded and is important for the amount of information that can be captured in the embeddings. However, larger embeddings also require more computational resources. Hence, we compare $1024$, $2560$, and $4096$-dimensional embeddings using embedding models of the same model family (see Section \ref{implementation}) to investigate the importance of embedding size in this use case. Note that the product information is structured as JSON before being embedded.

The parameter $top\_k$ determines how many candidates are retrieved in this step and either directly returned as search results or passed to the re-ranking step. A larger $top\_k$ increases the chances of finding relevant products but may also introduce more irrelevant products. Furthermore, it increases the computational cost and latency of the search, especially if re-ranking is applied to all retrieved candidates. Therefore, we evaluate the search performance and latency for $top\_k$ values of $20$, $50$, $100$, $150$, and $200$.

\textbf{Re-ranking:}
After retrieving the top $k$ candidates based on vector similarity, a re-ranker refines the initial candidate list by estimating the relevance of each candidate with respect to the query. Unlike the retriever, the re-ranker jointly processes the query and product information, enabling it to capture more complex semantic relationships. Although computationally more expensive, re-ranking has been shown to significantly improve retrieval performance \cite{ZhuLargeLanguage2025}. Therefore, we evaluate all configurations both with and without re-ranking.

\begin{figure}[htbp]
  \centering
  \def\svgwidth{0.7\linewidth}
  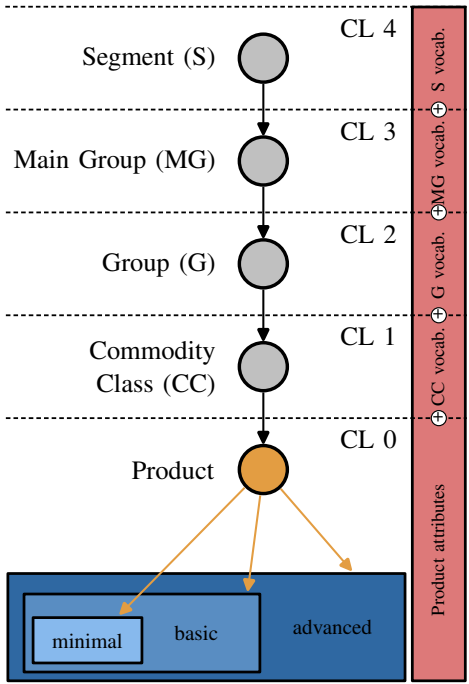
    \caption{Visualization of the ECLASS hierarchy and the corresponding product information for a specific product. A product is assigned to exactly one commodity class and can be represented with either the \textit{minimal}, \textit{basic}, or \textit{advanced} product data level. Additionally, the product information can be augmented with the name, definition, and keywords of different \aclp{cl} of the ECLASS hierarchy.}
\label{fig_database}
\end{figure}

\subsection{Incorporation of ECLASS vocabulary}
\label{incorporation}
We augment product embeddings with hierarchical ECLASS metadata by incrementally adding category name, definition, and keywords from higher levels of the taxonomy (CL1–CL4), while CL0 denotes product-only information. For instance, if a product is assigned to the commodity class \textit{"Measuring transformer disconnect terminal"}, we augment the product attributes with the metadata of this commodity class and refer to this as \acf{cl} $1$. For \ac{cl} $2$, we additionally include the information of the group \textit{"Terminal block systems"}, and so forth until the entire category path is covered as shown in Figure \ref{fig_database}. Consequently, we refer to the pure product information as \ac{cl} $0$.

This systematic evaluation reveals how hierarchical semantic context influences retrieval performance and identifies which category levels provide the most valuable information. These findings enable us to assess the semantic completeness and relevance of the ECLASS standard for product search applications, with implications for future standard development as discussed in Section \ref{discussion}.

\subsection{Dataset creation}
\label{dataset_creation}
To evaluate semantic product search in this industrial setting, we constructed a dataset of queries with corresponding relevant products. This is challenging in industrial domains due to domain-specific knowledge requirements, non-binary relevance judgments, and the risk of incomplete labeling in large catalogs. To address these challenges, we approached the dataset creation from both an expert and a non-expert perspective, aiming to cover a small set of high-quality samples as well as a more diverse collection of samples that may contain occasional mistakes or misconceptions.

For the expert perspective, domain experts from Phoenix Contact provided product search queries for control cabinet components and the corresponding relevant products based on their experience with customer inquiries. This resulted in $35$ samples with $1$ to $187$ and an average of $20.69$ ground truth relevant products per sample.

The non-expert perspective was covered by first-year trainees of Phoenix Contact by exploring product families and formulating search queries from a user perspective without relying on manufacturer-specific terminology. Therefore, they got a randomly sampled ground truth product from the database and should familiarize with its features and typical applications to create meaningful search queries across a wide range of product categories. This way, the trainees created $97$ additional evaluation samples, resulting in a total dataset of $132$ samples with an average of $29.93$ relevant products per sample. We refer to the different datasets as \textit{expert}, \textit{trainee}, and \textit{combined} dataset in the following.

To mitigate incomplete relevance labeling, we manually assessed samples in which at least once in $15$ representative experiments no relevant product was found in the top $10$ search results. This concerned $48$ queries for which we checked the top $10$ candidates in all $15$ experiments. For the representative experiments, we used an embedding size of $4096$, $top\_k = 200$, \textit{re-ranking}, no \textit{query rewriting}, and one of each following parameters:

\begin{itemize}
    \item \makebox[2.5cm][l]{category level:} $0$, $1$, $2$, $3$, $4$
    \item \makebox[2.5cm][l]{data level:} \textit{minimal}, \textit{basic}, \textit{advanced}
\end{itemize}
\noindent
Additionally, we reviewed the top $5$ candidates for all expert queries using the same experiment parameters to increase the reliability of evaluation metrics on the $5$ most relevant suggestions on this high-quality part of the dataset.

\section{Evaluation}
\label{evaluation}

\subsection{Implementation}
\label{implementation}
All components of the product search pipeline in our experiments are implemented using models from the \textit{Qwen3} model family \cite{qwen3technicalreport,ZhangQwen3Embedding2025}. This choice is motivated by several factors. First, it provides a unified ecosystem of models for different tasks, including text generation, embedding, and re-ranking, allowing for seamless integration and reducing incompatibilities between components. Second, \textit{Qwen3} models offer strong performance across multilingual and technical domains, which is particularly relevant for industrial product data characterized by domain-specific terminology and structured attributes \cite{qwen3technicalreport}. Third, the availability of models with varying sizes enables a systematic analysis of the trade-off between computational cost and performance. Finally, \textit{Qwen3} models are well-suited for local deployment, which is essential in industrial settings where data privacy, latency, and independence from external APIs are critical requirements.

For query rewriting, we employ \textit{Qwen3-8B} \cite{qwen3technicalreport}, which is used to transform natural-language queries into concise, attribute-focused representations. Product and query embeddings are generated using three variants of the \textit{Qwen3-Embedding} models \cite{ZhangQwen3Embedding2025} with different dimensionalities: \textit{Qwen3-Embedding-0.6B} ($1024$ dimensions), \textit{Qwen3-Embedding-4B} ($2560$ dimensions), and \textit{Qwen3-Embedding-8B} ($4096$ dimensions). This allows us to systematically evaluate the impact of embedding size on retrieval performance. For the re-ranking stage, we utilize \textit{Qwen3-Reranker-8B} \cite{ZhangQwen3Embedding2025}, which jointly processes query-candidate pairs to estimate relevance scores.

All experiments are conducted on a local server equipped with $8$ \textit{NVIDIA A100-SXM4} GPUs with $80\,GB$ of memory. The product database and vector store are implemented using Neo4j \cite{Neo4jGraphDatabase2026}. We leverage its \ac{ann} vector search capabilities to efficiently retrieve $top\_k$ candidates based on cosine similarity in the embedding space. This setup enables scalable and low-latency similarity search over the entire product database.

\subsection{Baselines}
\label{baselines}
To assess the effectiveness of the proposed approach, we compare it against both classical \ac{ir} methods and modern foundation model-based baselines.

As a traditional lexical baseline, we employ \textit{\ac{bm25}} \cite{RobertsonBM251994}, a probabilistic retrieval method that ranks documents based on term frequency and inverse document frequency, thereby relying on exact term matching between query and documents. In addition, we evaluate a \textit{\ac{bm25}} $+$ \textit{re-ranking} variant, where the initial \textit{\ac{bm25}} candidates are further refined using the same re-ranking model as in our dense retrieval pipeline. We retrieve the top $200$ candidates to ensure comparability with our best performing dense retrieval configuration and use the \textit{advanced} product data level to maximize the amount of information available for lexical matching.

To also benchmark against recent end-to-end \ac{ai}-based approaches, we include foundation model baselines with web search capabilities. Specifically, we evaluate \textit{Claude Sonnet~4.6} \cite{Claude46}, \textit{GPT-4.1} \cite{OpenAIGPT41}, and \textit{GPT-5.2} \cite{OpenAIGPT52}, which are prompted to recommend up to five relevant products for each query based on web searches restricted to the Phoenix Contact website. To ensure a fair comparison in terms of computational cost and token usage, each model is limited to a maximum of three web search calls per query. However, we found that the models reached this limit only three times in total. Since the website provides access to the entire product catalog, we manually assess candidates that are not contained in our database for the \textit{expert} dataset. The results are reported in Table \ref{tab_ablation_results}.

\subsection{Evaluation metrics}
\label{evaluation_metrics}
We evaluate the presented approaches using standard \ac{ir} metrics, focusing on the top-ranked results:

\begin{itemize}
    \item \textbf{Hit\_Rate@k:} fraction of queries for which at least one relevant product appears in the top $k$ results
    \item \textbf{Hits@k:} average number of relevant products within the top $k$ results
    \item \textbf{\ac{mrr}:} average reciprocal rank of the first relevant result
    \item \textbf{Recall:} fraction of all relevant products that are retrieved
    \item \textbf{R-Precision:} precision at rank $R$, where $R$ is the total number of relevant products for a given query
    \item \textbf{Precision@k:} fraction of relevant products among the top $k$ results
\end{itemize}

\noindent
In particular, we include a combination of \textit{R-Precision} and \textit{Precision@k}, which we refer to as \textit{R-Precision@k}. It is defined as follows:
\begin{equation*}
    \text{R-Precision@k} = \max\left(\text{R-Precision}, \text{Precision@k}\right)
\end{equation*}
\noindent
This metric extends standard \textit{R-Precision} by incorporating a cutoff $k$, ensuring meaningful evaluation when the number of relevant items exceeds the inspected result set. In such cases, classical \textit{R-Precision} may underestimate performance, whereas the combined metric captures both completeness and early precision.

This selection of metrics emphasizes the quality of the top-ranked candidates rather than exhaustive retrieval. In practical applications, both human users and \ac{llm}-based agents typically consider only a small number of results when searching for products. Therefore, it is more important that the first few returned items are highly relevant than that all possible relevant products are retrieved.

\renewcommand{\arraystretch}{1.7}
\begin{table}[htbp]
    \caption{Performance of the product data levels \textit{minimal}, \textit{basic}, and \textit{advanced} on the \textit{expert} dataset with and without \textit{re-ranking}. For these experiments, we used \ac{cl} $1$, embeddings size $4096$, $top\_k = 200$, and no \textit{query rewriting}. The values in parentheses represent the results on the dataset with manual assessment of the top candidates (see Section \ref{dataset_creation}).}
    \centering
    {\setlength{\tabcolsep}{5pt}
    \begin{tabular}{@{}|l|c|c|c|@{}}
        \hline
        \textbf{Experiment} & \textbf{Hit\_Rate@5}& \textbf{MRR}& \textbf{Recall} \\
        \hline
        minimal & $60.0\% \left(74.3\%\right)$ & $47.3\% \left(62.2\%\right)$ & $76.3\% \left(78.9\%\right)$ \\
        basic & $62.9\% \left(74.3\%\right)$ & $55.1\% \left(67.8\%\right)$ & $\mathbf{77.2\% \left(80.7\%\right)}$ \\
        advanced & $68.6\% \left(82.9\%\right)$ & $50.0\% \left(64.9\%\right)$ & $74.3\% \left(74.8\%\right)$ \\
        \hline
        minimal\textsubscript{rr} & $77.1\% \left(88.6\%\right)$ & $67.8\% \left(84.6\%\right)$ & $76.3\% \left(78.9\%\right)$ \\
        \textbf{basic\textsubscript{rr}} & $\mathbf{82.9\% \left(94.3\%\right)}$ & $\mathbf{70.9\% \left(87.4\%\right)}$ & $\mathbf{77.2\% \left(80.7\%\right)}$ \\
        advanced\textsubscript{rr} & $77.1\% \left(85.7\%\right)$ & $60.9\% \left(76.6\%\right)$ & $74.3\% \left(74.8\%\right)$ \\
        \hline
    \end{tabular}}
    \label{tab_main_results}
\end{table}
\renewcommand{\arraystretch}{1}

\section{Results and Discussion}
\label{results}

\subsection{Main results}
\label{main_results}
Our experiments demonstrate that the combination of structured product descriptions with dense vector retrieval and \textit{re-ranking} achieves strong performance on semantic product search for electronic components (Table \ref{tab_main_results}). The \textit{basic} data level combined with \textit{re-ranking}, denoted as basic\textsubscript{rr}, yields the best results with a Hit\_Rate@5 of $82.9\%$ on the original \textit{expert} dataset, which increases to $94.3\%$ when evaluated against the manually assessed candidates. This represents a significant improvement over approaches using only dense retrieval without \textit{re-ranking}, where the same configuration achieves $74.3\%$ Hit\_Rate@5. A similar trend can be observed across all product data levels, where \textit{re-ranking} consistently improves the \ac{mrr} by approximately $10$--$20\%$, confirming the importance of this component in the retrieval pipeline.

The \textit{basic} data level demonstrates the most effective balance between information richness and retrieval performance (Table~\ref{tab_main_results}). While the \textit{advanced} level includes more attributes, it yields lower scores across all metrics compared to \textit{basic}, suggesting that excessive attribute information may introduce noise or redundancy that interferes with vector similarity matching. Conversely, the \textit{minimal} level performs in between the two alternatives, especially when \textit{re-ranking} is applied, indicating that most of the relevant information for our evaluation samples is already captured in the manufacturer-specific short descriptions, but the inclusion of key attributes in the \textit{basic} level provides a significant boost in performance.

\begin{figure}[htbp]
\centerline{\includegraphics[width=\linewidth]{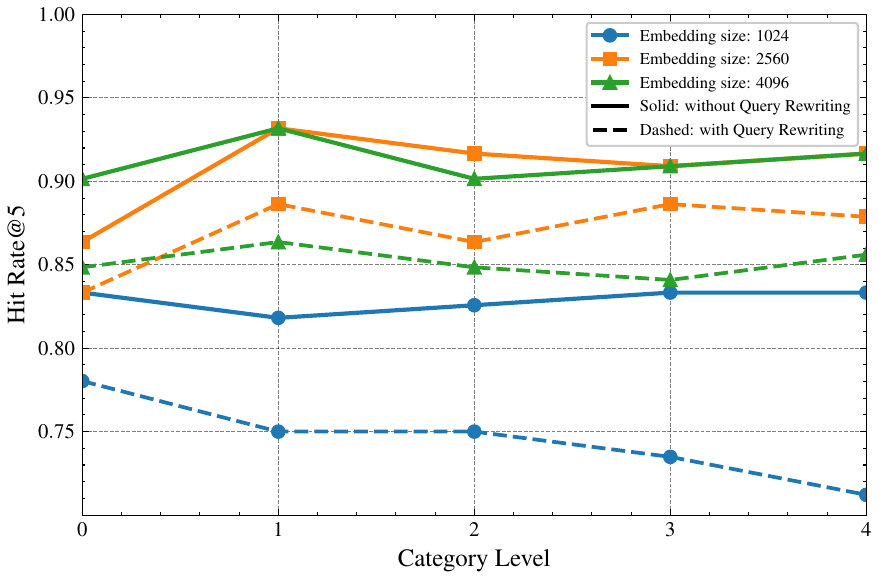}}
\caption{Search performance over different \acp{cl} and embedding sizes with and without \textit{query rewriting} on the \textit{combined} dataset with manual assessment of the top candidates (see Section \ref{dataset_creation}). For these experiments, we used the \textit{basic} product data level, \textit{re-ranking}, and $top\_k = 200$.}
\label{fig_category_levels_qr}
\end{figure}

Regarding the impact of the ECLASS hierarchy on search performance, Figure \ref{fig_category_levels_qr} shows that augmenting product embeddings with category metadata (\acp{cl} $1$--$4$) tends to improve the performance compared to using only product information (\ac{cl} $0$). The improvement is more pronounced with \textit{re-ranking} enabled and with \ac{cl} $1$, where the best configuration achieves the $93.2\%$ Hit\_Rate@5 on the \textit{combined} dataset. Embedding dimensionality also affects performance, with higher-dimensional embeddings ($2560$ and $4096$ dimensions) generally outperforming lower-dimensional ($1024$ dimensions). Query rewriting, however, shows a negative impact in all configurations, suggesting that our rewriting strategy may remove important information from the query.

The $top\_k$ parameter analysis (Figure \ref{fig_top_k_latency}) reveals the relationship between number of retrieved candidates and both performance and latency. Increasing $top\_k$ from $20$ to $200$ improves R-Precision@20 from $53.5\%$ to $68.9\%$ when re-ranking is applied, at the cost of linearly increasing latency. The mean latency grows from $1.7$ seconds at $top\_k = 20$ to $11.45$ seconds at $top\_k = 200$, which corresponds to our basic\textsubscript{rr} experiment in the results tables.

\begin{figure}[htbp]
\centerline{\includegraphics[width=\linewidth]{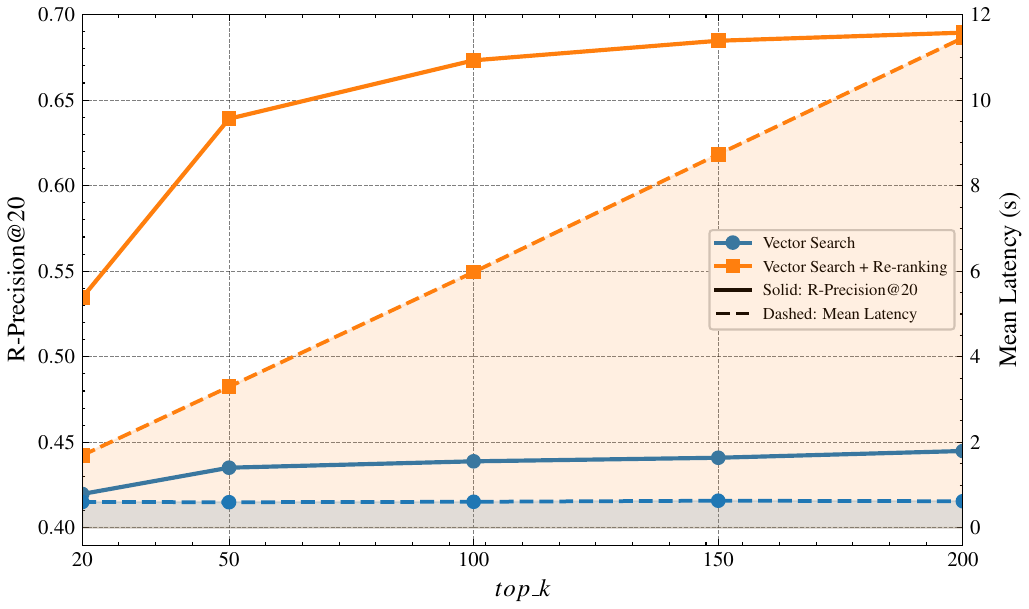}}
\caption{Search performance with and without \textit{re-ranking} over different values of $top\_k$ and the corresponding mean latency on the \textit{combined} dataset with manual assessment of the top candidates (see Section \ref{dataset_creation}). For these experiments, \ac{cl} $1$, \textit{basic} product data level, embeddings size $4096$, and no \textit{query rewriting}.}
\label{fig_top_k_latency}
\end{figure}

\renewcommand{\arraystretch}{1.6}
\begin{table*}[htbp]
    \caption{Performance comparison of our best configuration \textit{basic\textsubscript{rr}} with the full-text baseline \textit{\ac{bm25}} and the foundation model with web-search baselines. For \textit{\ac{bm25}} the \textit{advanced} product data level was used, to ensure that all available product information is included in the search. For the foundation model baselines the models were prompted to recommend up to $5$ products for each query based on a web-search on the Phoenix Contact website. This limits their achievable Recall performance indicated by *. Values are reported as \textit{expert/trainee} (E/T) to show dataset-specific behavior and quality differences.}
    \centering
    {\setlength{\tabcolsep}{4pt}
    \begin{tabular}{@{}|l|c|c|c|c|c|c|c|@{}}
         \hline
            \textbf{Experiment} & \textbf{Hit\_Rate@5 (E/T)} & \textbf{Hits@5 (E/T)} & \textbf{MRR (E/T)} & \textbf{Recall (E/T)} & \textbf{R-Precision@5 (E/T)} & \textbf{Latency (s)} & \textbf{Ext. Costs (\$)} \\
            \hline
            advanced\_bm25 & $31.4\%\,/\,59.8\%$ & $0.86\,/\,1.41$ & $24.2\%\,/\,46.1\%$ & $59.0\%\,/\,73.8\%$ & $26.1\%\,/\,41.7\%$ & $\mathbf{0.30}$ & $\mathbf{0.00}$ \\
            advanced\_bm25\textsubscript{rr} & $85.7\%\,/\,88.7\%$ & $2.69\,/\,2.49$ & $73.7\%\,/\,74.3\%$ & $59.0\%\,/\,73.8\%$ & $59.3\%\,/\,66.1\%$ & $13.06$ & $\mathbf{0.00}$ \\
            \hline
            Claude Sonnet~4.6 \cite{Claude46} & $91.4\%\,/\,-$ & $\mathbf{3.51}\,/\,-$ & $\mathbf{90.0\%}\,/\,-$ & $31.1\%^*\,/\,-$ & $73.1\%\,/\,-$ & $49.41$ & $0.29$ \\
            GPT-4.1 \cite{OpenAIGPT41} & $82.9\%\,/\,-$ & $2.17\,/\,-$ & $76.9\%\,/\,-$ & $16.7\%^*\,/\,-$ & $43.7\%\,/\,-$ & $16.42$ & $0.04$ \\
            GPT-5.2 \cite{OpenAIGPT52} & $71.4\%\,/\,-$ & $2.43\,/\,-$ & $71.4\%\,/\,-$ & $17.9\%^*\,/\,-$ & $49.2\%\,/\,-$ & $45.51$ & $0.49$ \\
            \hline
            basic\textsubscript{rr} (ours) & $\mathbf{94.3\%}\,/\,\mathbf{92.8\%}$ & $\mathbf{3.43}\,/\,\mathbf{2.87}$ & $\mathbf{87.4\%}\,/\,\mathbf{81.7\%}$ & $\mathbf{80.7\%}\,/\,\mathbf{83.4\%}$ & $\mathbf{76.0\%}\,/\,\mathbf{74.2\%}$ & $\mathbf{11.45}$ & $\mathbf{0.00}$ \\
            \hline
    \end{tabular}}
    \label{tab_ablation_results}
\end{table*}
\renewcommand{\arraystretch}{1}

\subsection{Baseline comparison and practical trade-offs}
\label{ablations}
To contextualize the main findings, we compare our best dense retrieval configuration basic\textsubscript{rr} against lexical and foundation model baselines in Table \ref{tab_ablation_results}. Across both the \textit{expert} and \textit{trainee} subsets, basic\textsubscript{rr} achieves the strongest overall balance of top-ranked relevance and coverage. In particular, it reaches the highest Hit\_Rate@5, Recall, and R-Precision@5 among all evaluated methods on both subsets, indicating that the approach is robust not only for high-quality expert queries but also for broader trainee-generated queries.

The \textit{\ac{bm25}} baseline shows a clear lexical mismatch effect, especially on the \textit{expert} subset, where manufacturer-specific wording differs from user phrasing. Adding instruction-aware re-ranking to \textit{\ac{bm25}} strongly improves top $5$ results, e.g., Hit\_Rate@5, but Recall remains bounded by the initial lexical candidate set. This confirms that \textit{re-ranking} is highly beneficial but cannot compensate for missing semantically relevant candidates in the first-stage retrieval.

For the foundation model web-search baselines, the values in Table \ref{tab_ablation_results} are only reported for the \textit{expert} subset because of the manual evaluation effort. When interpreting Recall for these baselines, it is important to note that they are prompted to return at most five products per query. For queries with more than five relevant products, this output cap inherently limits achievable coverage compared to retrieval pipelines that can return larger candidate sets.

Within this setup, \textit{GPT-4.1} and \textit{GPT-5.2} perform significantly worse than basic\textsubscript{rr} across all reported metrics, while \textit{Claude Sonnet~4.6} achieves reasonably strong top $5$ search performance. However, this comes with substantially higher latency and costs. In our experiments, the average latency for \textit{Claude Sonnet~4.6} is $49.41$ seconds per query, which is more than four times higher than basic\textsubscript{rr} with $11.45$ seconds. Additionally, the per-query cost for \textit{Claude Sonnet~4.6} is approximately $0.29\,\mathdollar$, which can quickly accumulate in large-scale applications, while basic\textsubscript{rr} incurs zero external costs in our local deployment setup.

Beyond these efficiency trade-offs, reliability is an additional concern for foundation model baselines. Among the evaluated models, \textit{GPT-4.1} produced $9$ hallucinated products that do not exist in the catalog across the $35$ expert queries, which is more problematic in practice than returning irrelevant but existing products. Taken together, these findings indicate that the proposed pipeline is the more reliable and cost-efficient choice for industrial semantic product retrieval.

\subsection{Qualitative analysis and discussion}
\label{discussion}
The quantitative results indicate that our best configuration basic\textsubscript{rr} is highly effective for practical use, but qualitative error analysis reveals remaining failure modes that are relevant for deployment. On the \textit{expert} dataset, a Hit\_Rate@5 of $94.3\%$ implies that for $2$ of $35$ samples no relevant product appears in the top 5. The two critical queries are:
\begin{enumerate}
    \item \textit{"Which terminal blocks have the highest connection density per terminal block?"}
    \item \textit{"Which Snap-in terminal blocks can I use?"}
\end{enumerate}
\noindent
These examples expose two distinct limitations. First, the \textit{"connection density"} query requires reasoning over a derived technical criterion rather than matching an explicitly stored attribute. Pure dense retrieval with re-ranking is not designed to reliably compute such aggregate or ratio-like characteristics from heterogeneous product fields. Second, the \textit{"Snap-in"} query illustrates terminology ambiguity in a highly specialized domain. In our dataset, the intended target products use \textit{push-x} connection technology, whereas the query term is semantically closer to \textit{push-in}. Because both technologies are present and lexically related, the retrieval pipeline tends to rank \textit{push-in} products ahead of the intended \textit{push-x} alternatives.

Beyond failure analysis, qualitative inspection also clarifies why ECLASS enrichment improves retrieval quality. Consider the query \textit{"Which terminal blocks with longitudinal disconnect features for transformer measurement are available for conductor sizes from $1.5$ to $6\,mm^2$, equipped individually with slotted or cross-slotted screws?"}. We found that the transformer measurement context is not explicitly encoded in the product short descriptions or attributes of the relevant products, but it is captured by the assigned ECLASS commodity class \textit{"$27-25-01-09$ Measuring transformer disconnect terminal"} and its definition \textit{"Terminal blocks with longitudinal and transverse disconnect function for the connection of current and voltage transformers"}. Adding this hierarchical semantic context to product representations enables relevant products to be retrieved that would otherwise remain difficult to identify.

Taken together with Figure \ref{fig_category_levels_qr}, this qualitative evidence supports the interpretation that ECLASS metadata acts as a semantic bridge between user intent and sparse product-internal wording. However, the analysis also highlights a structural limitation of the current standard for retrieval use cases. Although a general definition of each category is intended, ECLASS~13.0 only provides definitions for $68.18\%$ and keywords for $52.84\%$ of the considered categories. This incompleteness directly constrains the potential gains from hierarchy-aware embeddings. Therefore, an important practical implication is that retrieval quality is influenced not only by model selection and pipeline design, but also by the semantic completeness of international description standards.

From a practical perspective, these insights suggest that further improving industrial semantic search should be addressed from different angles. First, finding products that meet derived technical criteria requires advanced retrieval strategies that go beyond pure semantic similarity. This could involve hybrid retrieval approaches that combine vector similarity search with explicit handling of calculation-based attributes. Second, disambiguating closely related domain terms, such as \textit{snap-in} vs. \textit{push-in}, may require terminology normalization or synonym control mechanisms to ensure that user intent is correctly mapped to product features. Third, future releases of international description standards like ECLASS should consider the requirements of semantic applications and aim to enrich their vocabulary accordingly, especially in areas where product metadata is sparse. Addressing these aspects is likely to further improve the performance and robustness of industrial semantic product search, especially for more complex queries.

\section{Conclusion}
\label{conclusion}
This paper presented a systematic evaluation of \ac{llm}-assisted dense retrieval for semantic product search on structured industrial component data using ECLASS as an international description standard. Across the studied pipeline components, we found that combining \textit{basic} product data with semantic enrichment of the ECLASS \textit{commodity classes} and re-ranking provides the most effective configuration. On the manually assessed \textit{expert} dataset, this setup achieves a Hit\_Rate@5 of $94.3\%$ showing the potential of established description standards to enhance industrial semantic product search by providing structured hierarchical context and semantic information that complements attribute-centric product descriptions. 

Compared with lexical and foundation model web-search baselines, the proposed approach offers the best overall balance of search performance and efficiency for our use case, while enabling local deployment without external inference costs. At the same time, the qualitative analysis highlights remaining challenges for derived technical criteria and terminology ambiguity, as well as limitations from incomplete category metadata. Future work should therefore focus on addressing these challenges through advanced retrieval techniques, ambiguity resolution, and enrichment of description standards, to further enhance the capabilities of industrial semantic product search.

\section*{Acknowledgment}
The authors gratefully acknowledge Phoenix Contact GmbH \& Co. KG and PHOENIX CONTACT Deutschland GmbH for their expertise and support in creating the evaluation dataset.

\bibliographystyle{IEEEtran}
\bibliography{IEEEabrv,references}

\end{document}